# Pressure-Driven Activation of Semi-core *p*-Orbitals in Rb and Cs via Crystal-Field Splitting

Shuran Ma,[1] Shuyang Sun,[1] Dengyao Yu,[1] Xue Cong,[1] Yanchao Wang,[2]* Yuanzheng Chen,[3]* and Zhen Liu[1,4]*

[1] *School of Physics and Astronomy, Beijing Normal University, Beijing 100875, China*

[2] *State Key Laboratory of Superhard Materials and International Center of Computational Method & Software, College of Physics, Jilin University, Changchun 130012, China*

[3] *School of Physical Science and Technology, Southwest Jiaotong University, Chengdu 610031, China*

[4] *Key Laboratory of Multiscale Spin Physics (Ministry of Education), Beijing Normal University, Beijing 100875, China*

**Abstract:** High-pressure unconventional oxidation states are attributed to exceptional charge transfer, which is itself induced by the differential pressure response of atomic orbital energy levels. The high oxidation states observed in heavy alkali metals (Cs, Ba, Ra) exemplify this behavior, as their semi-core *p*-states surpass the valence orbital levels of oxidants, e.g., the 2*p*-orbital of F, at high pressure. However, semi-core *p*-electron activation in lighter alkali metals, such as Rb, is hindered because the relative orbitals lie at intrinsically lower energies due to their smaller radii, making pressure-induced upshift challenging. We demonstrate that crystal-field splitting of alkali-metal semi-core *p*-orbitals selectively elevates a portion of these states, substantially lowering the pressure threshold for their chemical activation. Here, our computations reveal that the semi-core 4*p*-electrons of Rb form weak metallic bonds with F in $P2_1/c$-$RbBF_5$ at around 250 GPa, a pressure accessible in modern high-pressure experiments. Extending to Cs-B-F systems, the crystal-field splitting drives 5*p* participation in Cs-F bond formation in $CsBF_5$. These findings unveil a new paradigm for semi-core or inner-shell chemistry, providing a unifying framework for semi-core electron activation and bonding. Crucially, they establish that semi-core electrons of light main-group elements, including Rb, are accessible at experimentally feasible pressures.

**Keywords:** Semi-core Electron; Crystal-field Splitting; High Pressure; Rubidium; Cesium.

## Introduction

The reduction of interatomic distances under compression can reorder electronic energy levels in solids,[1–4] giving rise to phenomena such as unconventional oxidation states,[5,6] unexpected structural transition,[7] and emergent novel optical and electronic properties.[8,9] A convenient way to rationalize these diverse behaviors is to start from the element-specific response of atomic orbital energies to confinement. Accordingly, a compressed-atom framework compares the relative orbital energies of isolated atoms under spherical confinement to interpret pressure-driven electronic reordering.[3] In this picture, orbitals with a larger number of radial nodes undergo a more pronounced energy increase upon compression, leading to systematic reordering among *s*, *p*, and *d* levels. A prominent consequence is pressure-induced *s*-*d* transfer, which can make *s*-block metals exhibit transition-metal-like characteristics.[10–15] For example, Na has been reported to transform from a metal into a transparent insulating electride at high pressure, where pressure-enhanced *p*-*d* hybridization is believed to localize electrons in interstitial regions[8,16] More generally, inter-element orbital-energy comparisons within this framework provide a useful guideline for understanding unconventional charge transfer and bonding motifs under compression.[17–21] For instance, Cs 5*d* states can accept electrons from Li and render Cs anionic at high pressure,[22] consistent with the prediction that the compressed Cs 5*d* level drops below the Li 2*s* level.

Strikingly, the chemically inert semi-core *p*-electrons of alkali(-earth) elements can become reactive under compression when their energies are driven above the valence levels of strong oxidants (e.g., the F 2*p* states). For instance, Cs has been reported to form Cs-F bonds involving its semi-core 5*p* electrons at high pressure, as the Cs 5*p* level is globally pushed above the F 2*p* level, enabling oxidation states as high as +5.[23] Consistently, Cs was also predicted to attain high oxidation states in Cs-O compounds above 221 GPa.[24] Similar semi-core electron participation has been proposed for Ba, where the 5*p* electrons become oxidizable under high pressure.[25] Extending this trend to the next period, Ra is predicted to reach oxidation states up to +8 in $RaF_n$ stoichiometries, with its inner 6*p* shell becoming chemically active under compression.[26]

In contrast to heavy alkali and alkaline-earth fluorides, the chemical activation of semi-core *p*-electrons in lighter alkali metals, such as Rb, is intrinsically more challenging due to their smaller radii and lower orbital energies. Indeed, it has been reported that in Rb-F binaries Rb is unlikely to be oxidized beyond the +1 state via 4*p* activation even at very high pressure.[27]

This reduced propensity can be rationalized within the atomic-compression picture. Compared with heavier analogues (Cs, Ba, Ra), the Rb 4*p* orbital has fewer radial nodes and therefore undergoes a smaller pressure-induced energy upshift, making the requisite interatomic level crossing with oxidant-derived states less attainable. Moreover, the Rb 4*p* level remains below the F 2*p* manifold over a broad pressure range, further disfavoring charge transfer under compression. Whether inner-shell activation chemistry can be systematically extended to lighter elements thus remains an open question.

In this study, however, our first-principles calculations reveal that the Rb 4*p* semi-core electrons participate in bonding with F in the metastable ternary $RbBF_5$ at 200-300 GPa, which lies beyond the predictive scope of simple isolated-atom energy-level comparisons under compression. This phase adopts a layered architecture in which two-dimensional Rb-F sheets (Rb:F = 1:1) alternate with layers of $[BF_4]^-$ units in an ABAB… stacking sequence. Each Rb atom resides in a 12-fold coordination environment, with the F atoms occupying the vertices of a truncated-cube-like polyhedron. This geometry breaks local symmetry, as the Rb $4p_x$, $4p_y$, and $4p_z$ orbitals are oriented toward crystallographically inequivalent F sites. Consequently, a pronounced crystal-field splitting of the Rb 4*p* manifold into higher-energy in-plane $4p_{x,y}$ and lower-energy out-of-plane $4p_z$ components. The $4p_{x,y}$ levels are lifted toward the F $2p_{x,y}$ states by the splitting, enabling this semi-core electron participation and promoting metallic Rb-F bonding within the layers. Furthermore, we also examine the effect of crystal-field splitting on activation of the semi-core 5*p* in Cs at high pressure. The structure of $CsBF_5$ consists of Cs-F chains and $[BF_4]^-$ units. The Cs-F chains are characterized by metallic bonding between alternating Cs and F atoms, attributed to crystal-field splitting that lifts the energy of one specific suborbital of the global 5*p* manifold, thereby promoting its interaction with F. Additionally, we observe covalent $[CsF_2]^+$ units in $CsBF_6$, which are a typical $sp^3d$ hybdridization and isoelectronic to $XeF_2$.[23,28] The formation of these Cs-F covalent bonds indicates that the elevation of the entire 5*p* manifold governs the semi-core 5*p* electron activation in this system instead of crystal-field splitting. It should be noted that crystal-field splitting of *p*-orbitals has received limited attention, since they tend to form covalent bonds via hybridization, and their splitting is often subtle in octahedral or cubic coordination environments under ambient conditions. However, our findings reveal that crystal-field splitting in asymmetric coordination environments lifts the degeneracy of semi-core *p*-orbitals in alkali metals such as Rb, selectively elevating suborbital energy levels and thereby promoting their chemical activation at high pressures.

## Computational Methods

**Structure Search** The package CALYPSO (Crystal structure AnaLYsis by Particle Swarm Optimization),[29–31] implemented with particle swarm optimization (PSO), is employed to predict the low enthalpy candidate structures from the ternary Cs-B-F and Rb-B-F stoichiometries. The techniques such as bond characterization matrice, and coordinate characterization functions are implemented in this package for the enhancement of prediction accuracy and efficiency. In this work, we use this package to predict and design the structure for $CsBF_n$ and $RbBF_n$, where $n = 4, 5, 6, 7$, at the pressures of 100 GPa, 200 GPa, and 300 GPa, respectively. In the procedure of prediction, 40% of structures in each iterated generation are randomly generated structure, while 60% structures are selected from previous generation.

**Formation Enthalpy and Electronic Structure** To access the thermal stability and electronic structure, the density functional theory (DFT) package Vienna Ab initio Simulation Package (VASP)[32] was employed to run the computation for the readily searched $CsBF_n$ and $RbBF_n$ structures. Pseudopotential method is widely used in density functional theory for the description of the interaction between valence electrons and core charge. In this work, the pseudopotential is generated by the framework of projector augmented wave (PAW) method.[33,34] The exchange-correlation functional is considered within the generalized gradient approximation under the framework of Perdew-Burke-Ernzerhof.[35] In our calculations, the cut-off energy is set to be 1400 eV. Monkhorst-Pack mesh with a density of $2\pi \times 0.04$ Å$^{-1}$ are used to ensure the convergence.

**Ab Initio Molecular Dynamics** The explore the dynamic stability of $P2/c$-$RbBF_5$ beyond harmonic approximation, we apply the ab initio molecular dynamics (AIMD) for the relaxation of $P2/c$-$RbBF_5$ under NVT ensemble with Nosé-Hoover thermostat.[36–38] The computation runs with 0.5 fs for each ionic step and 20000 step in total. The conventional cell of $P2/c$-$RbBF_5$ was expanded into $2 \times 3 \times 2$ supercell with 168 atoms.

## Results

**Stability of Asymmetric Coordinating Structure** We have used first-principles PSO crystal-structure searching techniques to determine the stable phases of various compositions of $CsBF_n$ and $RbBF_n$ ($n$ = 4, 5, 7, 9). The thermal stability is accessed by ternary convex hull from the elemental formation enthalpy:

$$\Delta H^f = \frac{H[\mathrm{Cs(or\ Rb)BF}_n] - H[\mathrm{Cs\ (or\ Rb)}] - H[\mathrm{B}] - nH[\mathrm{F}]}{n+2}$$

The negative value suggests the stability against decomposition into the pure elemental substances. The vanishing of the energy above the convex hull signifies thermal stability. However, a large portion of compounds in database such as Materials Project[39] can be synthesizable even though the formation enthalpy lies with several tens of meV above the convex hull.[40–42] Here, the compoistions with value of the energy above hull lower than 60 meV are recognized as synthesizable metastable structures for the both cases of $CsBF_n$ and $RbBF_n$, respectively.

The triangle phase diagram in terms of the energy above hull of $CsBF_n$, as illuatrated in Figure 1a, led us to predicted three synthesizable stoichiometries, $CsBF_4$, $CsBF_5$, and $CsBF_6$. The structural parameters of these structure can be found in Supplementar Table S1-S3. Our computation supposes that $CsBF_4$ is stabilized from ambient condtion upto around 250 GPa, see Supplementary Figure S1. The first principle structure method shows this stoichiometry is stabilized in structure with space group of $Pna2_1$ between 0GPa and 58 GPa, and transforms into one with space group of *Pnma*[43] upon the pressure increasing. As shown in Figure 1b, the enthalpy of $CsBF_5$ lies no more than 60 meV per atom above that of $CsBF_4$+F for pressures above 29 GPa, suggesting that $CsBF_5$ may be synthesizable under pressure. Between 41 and 264 GPa, the $CsBF_4$+$CsBF_6$ combination exhibits the lowest enthalpy. At 300 GPa, $CsF_2$+$BF_3$ becomes slightly more favorable, lying about 5 meV per atom below $CsBF_4$+$CsBF_6$. Notably, the enthalpy of $CsBF_5$ is only around 1 meV per atom higher than that of $CsBF_4$+$CsBF_6$ at 100 GPa. Across the 50-300 GPa range, the enthalpy difference between $CsBF_5$ and $CsBF_4$+$CsBF_6$ remains within 20 meV per atom.

Our structural search and first principle computation predict $CsBF_5$ as a potentially synthesizable metastable phase at high pressure. The candidate $P2_12_12$ structure, see Supplementary Figure S2a, is built from Cs-F chains and $BF_4$ molecular units. However, its phonon spectrum exhibits imaginary frequencies at 100 GPa along the Γ-X-S-Y-Γ and Z-U-R-R-Z paths revealing its dynamic instability, details of which can be found in Supplementary Figure S2 b and c. This indicates that the structure corresponds to a saddle point on the potential energy surface, implying a spontaneous distortion toward a dynamically stable configuration.[44] Starting from the candidate structure, we performed structural relaxation using expanded supercells and random atomic displacements (details in Supplementary Figure S3), which yielded a lower-enthalpy $P2_1/c$ polymorph as shown in Figure 1 e and f, which retains the Cs-F chain and $BF_4$ motif, but with a relative sliding of the chains and rotation of the $BF_4$ units. At 100 GPa, the Cs-F distances along the chain alternate between 2.201 and 2.114 Å. These are notably shorter than the Cs-F bond in B2-CsF (2.545 Å at the same pressure) and close to the Cs-F distance in $CsF_2$ (2.121 Å, from our re-calculation based on the parameters of ref[23]), hinting that inner 5*p* electrons may be involved in the bonding of $P2_1/c$-$CsBF_5$. The coordination of Cs in $P2_1/c$-$CsBF_5$, see Figure 1f, involves two F atoms along the chain and several crystallographically inequivalent F atoms from $BF_4$ units, resulting in a reduced local symmetry. It will be explicit in the following sections that semi-core 5*p* orbital undergoes splitting under the low-symmetry crystal field at high pressure, which is responsible for its chemical activation. Besides, Its dynamic stability is corroborated by phonon dispersion calculations, see Supplementary Figure S4.

The phase behavior of $CsBF_6$ is summarized in Figure 1c. Between 0 and 28 GPa, the compound is thermodynamically unstable with respect to $CsBF_4$ and elemental F. It becomes stable in a $P2_13$ structure from 28 to 68 GPa, and subsequently transforms to a $P3_121$ phase that remains stable up to 227 GPa. The $P3_121$ structure is built from alternating $CsF_2$ and $BF_4$ layers (Figure 1f). At 200 GPa, the Cs-F bond length within the $CsF_2$ unit is 1.853 Å, a value comparable to that in highly oxidized binary $[CsF_2]^+F^-$ at the same pressure.[23] This structural and metric similarity provides strong evidence for an unusually high oxidation state of Cs in this compound.

The ternary phase diagram of Rb-B-F propose the stability of two different synthesizable stoichiometries, $RbBF_4$ and $RBBF_5$, see Figure 2a. As the most stable stoichiometry, $RbBF_4$ adopts the prior synthesized *Pnma* structure[45] at ambient pressure and the $P4/n$ symmetry at above around 13 GPa based on our computation. For $RbBF_5$, its enthalpy locates above the convex hull within 60 meV per atom at 300 GPa. At ambient pressure, $RbBF_5$ with the space group of $P2_12_12_1$ is of the lowest enthalpy. Upon increasing pressure, the $P2/c$-$RbBF_5$ gains the lower enthalpy than the $P2_12_12_1$ structure. The structural parameters can be found in Supplementary Table S4 and S5. To access the possible decomposition route into binaries of $RbBF_5$ at high pressure, its enthalpy compared with the enthalpy sum of combinations of reactants are calculated as shown Figure 2b. When compared with the combination of $RbF_2$+$BF_3$, *P*2/*c*-$RbBF_5$ is more energetically favorable except at ambient condition and above 200 GPa. At 300 GPa, $RbF_2$+$BF_3$ is lower by 52.79 meV per atom than *P*2/*c*-$RbBF_5$. It is worth to note that $RbBF_4$+F is the lowest enthalpy reactant combination in our computation at ambient. With increasing pressure, the enthalpy of this reactant combination increases compared with *P*2/*c*-$RbBF_5$. At 300 GPa, the enthalpy sum of $RbBF_4$+F is lower than the one of *P*2/*c*-$RbBF_5$ by 22.05 meV per atom. *P*2/*c*-$RbBF_5$ is higher than the lowest enthalpic structures within 60 meV per atom between 200 GPa and 300 GPa. This refers that *P*2/c-$RbBF_5$ might be metastable at high pressure. The details of the thermal stability can be found in Supplementary Figure S5. The phonon bands of *P*2/c-$RbBF_5$ exhibit imaginary frequency no more than respectively 0.5 THz and 2.0 THz along the paths of Γ-Z and Γ-Y2, see Supplementary Figure S6..

Applying the same procedure used to identify the dynamically stable *P*$2_1$/*c*-$CsBF_5$ to *P*2/c-$RbBF_5$, however, the supercell calculations with random atomic displacements consistently returned back to the original *P*2/*c* structure, indicating that effects beyond the harmonic approximation may play a crucial role in its stabilization. To further examine the dynamic stability, we carried out AIMD simulations to compute root mean square deviation (RMSD) of particle coordinates and atomic trajectories. The results at 1000 K and 250 GPa, as illustrated in Figures 2e and f, confirm that ions remain localized around their equilibrium positions, demonstrating dynamical stability. The RMSD and atomic trajectories at other temperatures and pressures can be found in Supplementary Figure S7-S9.This behavior is reminiscent of other systems, e.g., b.c.c. iron under high pressure and high temperature, where harmonic phonon spectra exhibit imaginary frequencies while AIMD confirms a stable structure.[46]

**Electronic Transferring and Bonding Properties** Electron transfer was assessed via Bader charge analysis within the Quantum Theory of Atoms in Molecules (QTAIM) framework, where atomic charges are obtained by integrating the electron density over basins bounded by zero-flux surfaces.[47–49] As shown in Figure 3a, the Bader charges of Rb in $RbBF_4$ and Cs in $CsBF_4$ are both below the +1 threshold through the whole computed pressure range. In contrast, the charges for Rb in $RbBF_5$, and Cs in $CsBF_5$ and $CsBF_6$ all exceed +1 upon pressure increasing, indicating their inner *p*-electron transfer from the alkali metal to the surrounding atoms. As noted in the preceding discussion, all ternary systems examined in this work contain tetrahedral $BF_4$ molecular units, which result from $sp^3$ hybridization by accepting an electron from either Cs 6s or Rb 5s orbital, see Figure 3c. Hence, the $BF_4$ groups are anionic, denoted here as $[BF_4]^-$. Besides, the inner *p*-electrons are available for bonding with the F atoms that are not part of the $BF_4$ units in the structure.

The electron localization function (ELF) was computed to further investigate the topological distribution of electrons and the associated bonding characteristics. By definition, ELF approaches unity in regions of strong electron localization and tends to zero where electrons are essentially absent. As shown in Figure 3d, the ELF distribution on the (001) plane, which contains the Rb-F sheet in $P2_1/c$-$RbBF_5$, exhibits diffuse boundaries between Rb and F atoms. The ELF value in the interatomic region is approximately 0.21, suggesting weak metallic bonding between the two species. In Figure 3e, the ELF values along the Cs-F chain in $P2_1/c$-$CsBF_5$ alternate between 0.28 and 0.30, likewise indicating a weakly metallic character. In contrast, Figure 3f reveals a highly localized electron distribution, implying that $CsF_2$ exists as isolated molecular units embedded within the bulk matrix, with covalent Cs-F bonding in $P3_121$-$CsBF_6$.

To further quantify the bonding characteristics, we calculated the integrated crystal orbital Hamiltonian population (ICOHP) for each compound,[50] with more negative values corresponding to stronger bonding interactions, details see Table 1. For $P2_1/c$-$RbBF_5$, the ICOHP of the Rb-F pair is -0.930 eV per pair at 250 GPa. In $P2_1/c$-$CsBF_5$, the Cs-F bonds with lengths of 2.201 and 2.114 Å exhibit ICOHP values of -1.496 and -1.236 eV per pair, respectively. For comparison, the ICOHP of the Al-Al bond in face-centered cubic (f.c.c.) Al metal is computed as -1.11 eV per pair. These values indicate that the Rb-F and Cs-F bonds display bonding strengths comparable to those of typical *p*-block metals. In contrast, the Cs-F bond in $P3_121$-$CsBF_6$ gives a much more negative ICOHP of approximately -3.844 eV per pair, consistent with covalent character and in full agreement with the ELF analysis discussed above. $P3_121$-$CsBF_6$ could be regarded as $[CsF_2]^+[BF_4]^-$ due to the covalence in $CsF_2$ and $BF_4$ units.

## Discussion

**Rb-4*p* orbital level is lower than that of F-2*p*** Comparison of orbital energy levels across different elements serves as a fundamental tool for predicting the propensity for electron sharing at high pressure. To this end, we employed the He-lattice compression scheme[3] to track the pressure-induced shifts of the Cs-5*p*, Rb-4*p*, and K-3*p* orbitals with respect to the F-2*p* level over the 0-800 GPa range. Orbital energy level comparison between different elements is a prior method for prediction of electron sharing between these elements at high pressure. Here, with He-lattice compression method,[3] we compress these atoms between 0 GPa and 800 GPa to see the Cs-5*p* and Rb-4*p* as well as K-3*p* relative to the F-2*p* energy level. As illustrated in Figure 4, Cs-5*p* orbital energy level is lower than F-2*p* at ambient pressure, and would cross over F-2*p* level at around 36 GPa, which agrees with the high oxidation states of Cs[23] at high pressrue. The Rb-4*p* level would not be higher than F-2*p* unless the pressure is higher than around 610 GPa, while K-3*p* level is lower than F-2*p* between 0 GPa and 800 GPa. These results indicate that the activation of Rb's *p*-orbitals would occur only at ultra-high pressures, whereas activating K's 3*p* orbitals remains extremely challenging even under extreme compression. Nevertheless, the inner 4*p* orbitals of Rb do participate in electron sharing with F in *P*2/*c*-$RbBF_5$ as discussed above. This suggests that pressure-induced activation of Rb's 4*p* electrons cannot be fully understood through a simple comparison of elemental energy levels alone.

**Activation of inner *p*-elelctron by Crystal Field Splitting** Intrinsically, the activation of Rb's semicore 4p electrons at pressures lower than predicted by the atomic-compression picture is driven by an overlooked effect, which is crystal-field splitting further lifting the suborbital of 4*p* level or Rb. As illustrated in Figure 5a, the global upshift of semicore p-orbitals alone may be insufficient to raise the light alkali metal's p-levels above those of the oxidant. However, when the alkali metal resides in a low-symmetry coordination environment, crystal-field splitting provides an additional energy lift, enabling the activation.

To unveil the mechanism of Rb's 4*p* electron activation by the crystal-field splitting, we employ the projected density of states (PDOS) of electron to analyze the energy level arrangement at high pressure. As depicted in Figure 5b, with the Rb-F sheet defined as the $xOy$ plane, each Rb center experiences a coordination environment comprising four in-plane F atoms and eight out-of-plane F atoms from $[BF_4]^-$ molecular units. In *P*2/*c*-$RbBF_5$, the PDOS shows that the $4p_x$ and $4p_y$ states of Rb are half-filled, whereas its $4p_z$ state is fully occupied. This low-symmetry ligand field lifts the degeneracy of the Rb 4*p* orbitals, producing two distinct groups of non-degenerate levels, which are $4p_x$/$4p_y$ and $4p_z$, a consequent phenomenon driven by crystal-field splitting effects. The splitting shifts the Rb $4p_x/4p_y$ states upward under pressure, approaching the F $2p$ level (Figure 5b), which aligns with the similar PDOS profiles and energy ranges of the two sets of orbitals. The crystal-field splitting effect is enhanced for $p$-orbitals under high pressure. The chemical activation of the Rb $4p$ orbital in $P2_1/c$-$RbBF_5$ at a relatively low pressure is driven by the compression-induced energy lifting arising from crystal-field splitting in low symmetry coordination.

Let us now examine the $5p$ electron activation in $P2_1/c$-$CsBF_5$ and $P3_121$-$CsBF_6$. The $P2_1/c$-$CsBF_5$ structure is built from Cs-F chains and $[BF_4]^-$ molecular units. As illustrated in Figure 5c, we define the $xOy$ plane such that the Cs-F chains extend along the $y$ direction. This low symmetry is expected to lift the degeneracy of the Cs $5p$ orbitals, selectively raising the $5p_y$ level toward the F $2p$ states, as reflected in the PDOS shown in Figure 5c, which is a signature of crystal-field splitting driven inner electron activation. The pressure-induced upshift of the entire $5p$ manifold is also observed in the Cs-B-F ternary systems. The PDOS in Figure 5d reveals no significant energy difference among the $5p_x$, $5p_y$, and $5p_z$ states of Cs below the Fermi level, indicating that the splitting in this system is negligible in this case. In $P3_121$-$CsBF_6$, $[CsF_2]^-$ molecular units are present, forming a covalent system analogous to $XeF_2$. The formation of these molecules is primarily driven by compression-induced elevation of the whole $5p$ orbital level for the $sp^3d$ hybridization, as discussed in a previous study.[23] We see the crystal-field splitting and global level elevation of the semi-core $5p$ orbital are competing factor for chemical activation.

## Conclusions

In this work, we predict the inner $4p$-electron of Rb could be chemically activated and form bonds with F no more 300 GPa in a metastable $P2/c$-$RbBF_5$ through computation method. The smaller radius of Rb than that of Cs, Ba and Ra might make its $4p$-electron activation rarely occurs even at extremely high pressure in Rb-F binary. However, the crystal-field splitting of semi-core $4p$ orbital of Rb would lift its $4p_x$ and $4p_y$ suborbitals close to F's $2p$ orbital in $P2/c$-$RbBF_5$ because of its low symmetric environment. Notably, this effect is beyond the atomic compression model. In the Cs-B-F ternary systems such as $CsBF_5$ and $CsBF_6$, we see both crystal-field splitting driven semi-core $5p$ electron activation and whole manifold raising induced hybridization for these two stoichiometries. Our work brings new aspects that the crystal-field splitting of semi-core orbitals is unexpected mechanism for their chemical activation at high pressure, which would pave a new way to search high oxidation for lighter *s*-block metal such as Rb. In addition, the semi- or inner- core electron bonding chemistry would be expanded to lighter elements.

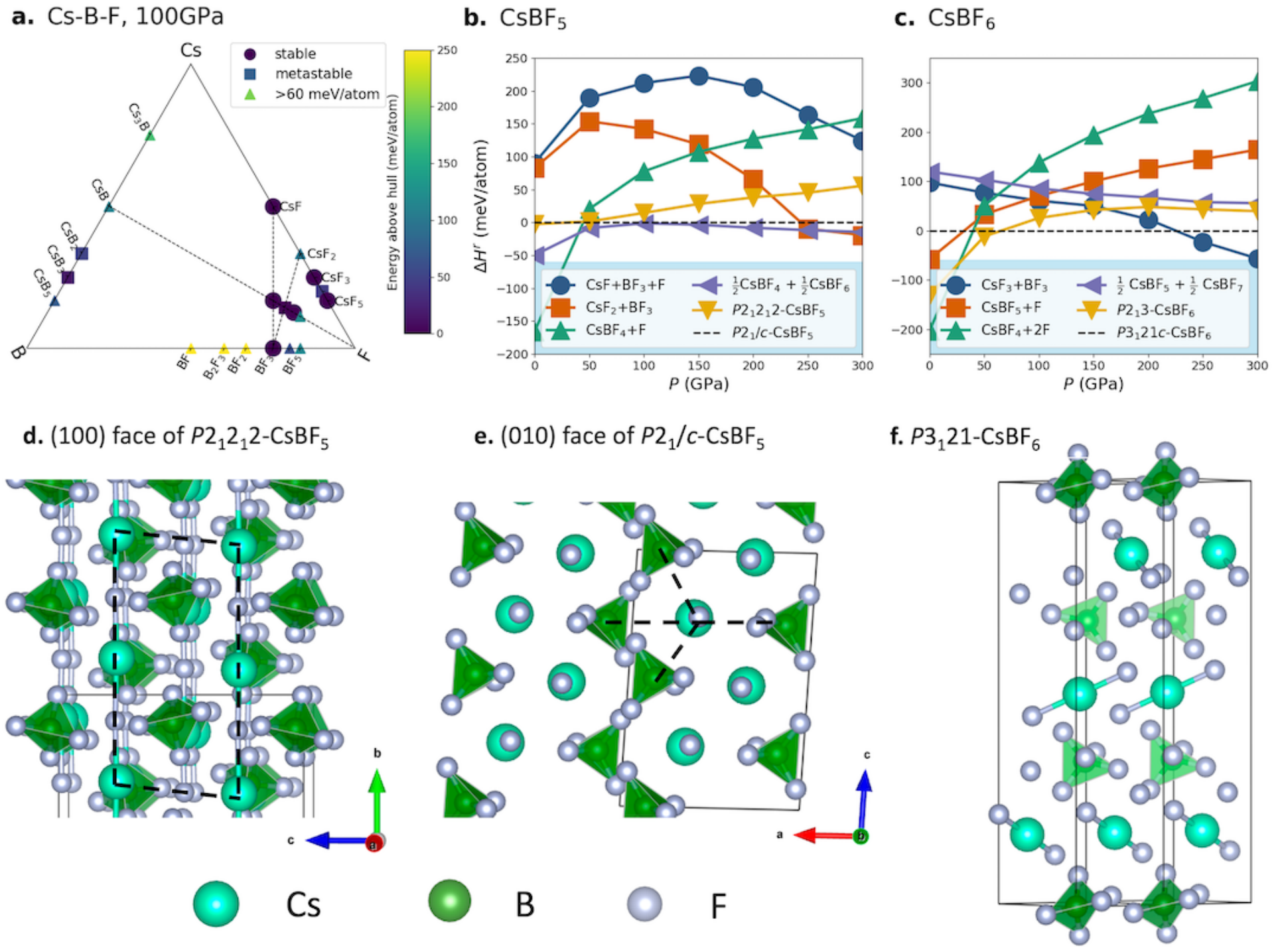

**Figure 1. The thermal stability and structure of Cs-B-F at high pressure. a.** The triangle phase diagram in term of the energy above hull at 100 GPa. The stoichiometries are classified into three categories, namely the stable, metastable and unstable. The stable stoichiometries refer as the ones whose formation enthalpy lying on the convex hull. The formation enthalpy of the metastable is higher than the convex hull within 60 emV per atom. The ones beyond this value are regards as unstable structures. **b** and **c** are the enthalpy comparation between composite $CsBF_{5,6}$ and the possible separate reactants combinations. **d** and **e** are the view perpendicular to (100) and (010) face of the structure $P2_1/c$-$CsBF_5$ at 100 GPa. In **e**,The relative slide between two neighbor Cs-F chains are highlighted. **f.** the structure of $P3_121$-$CsBF_6$ at 100 GPa.

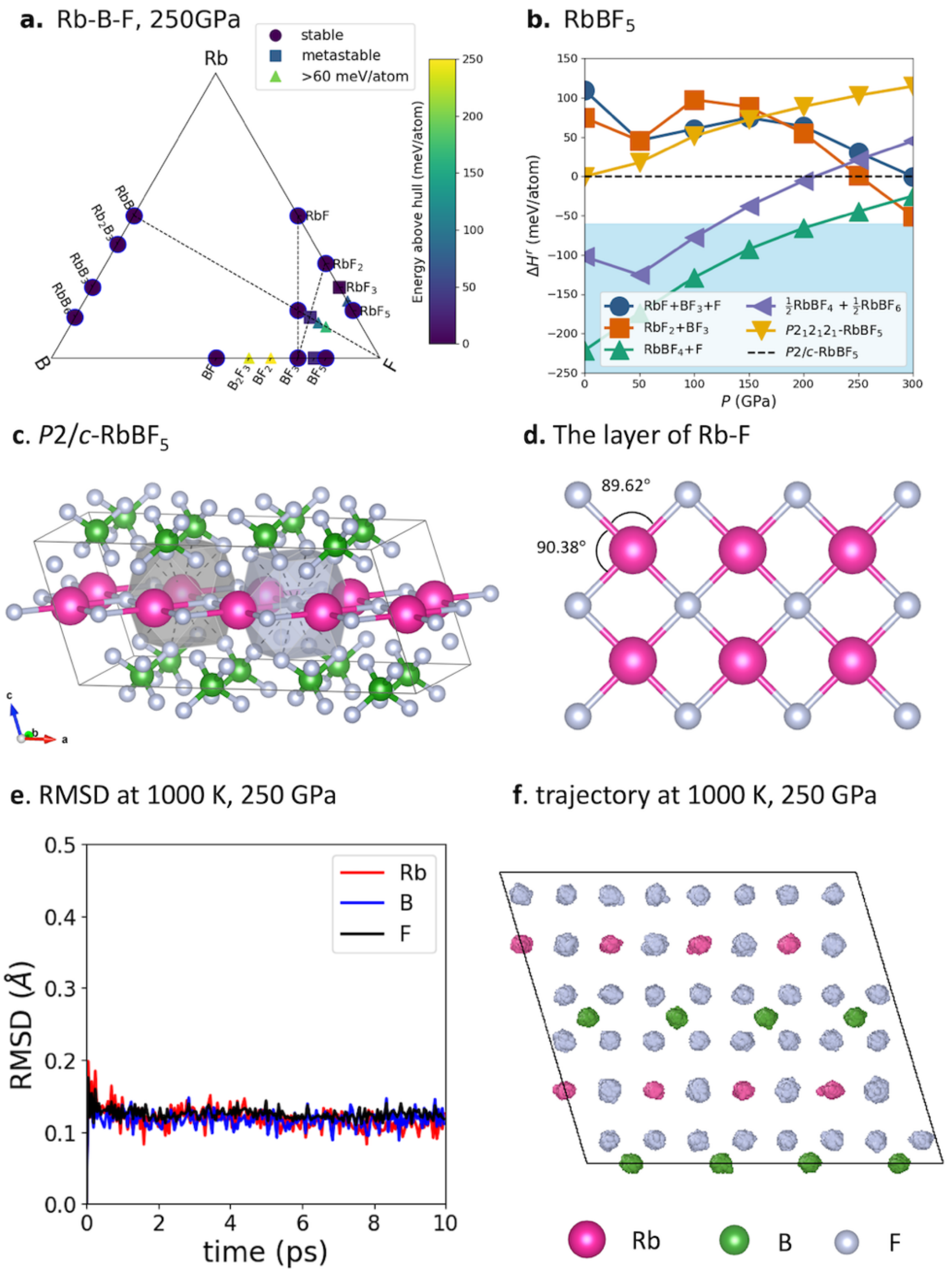

**Figure 2**. **The thermal stability and structure of Rb-B-F at high pressure. a.** The triangle diagram in terms of the value of the formation enthalpy above convex hull for Rb-B-F at 250 GPa. **b.** The enthalpy comparison between *P*2/c-$RbBF_5$ and selected synthetic routes. **c.** the layer structure of *P*2/c-$RbBF_5$ at 250 GPa. **d.** The top view of Rb-F two-dimensional sheet. **e** and **f** are the RMSD and trajectory of *P*2/c-$RbBF_5$ at 1000 K and 250 GPa computed by AIMD.

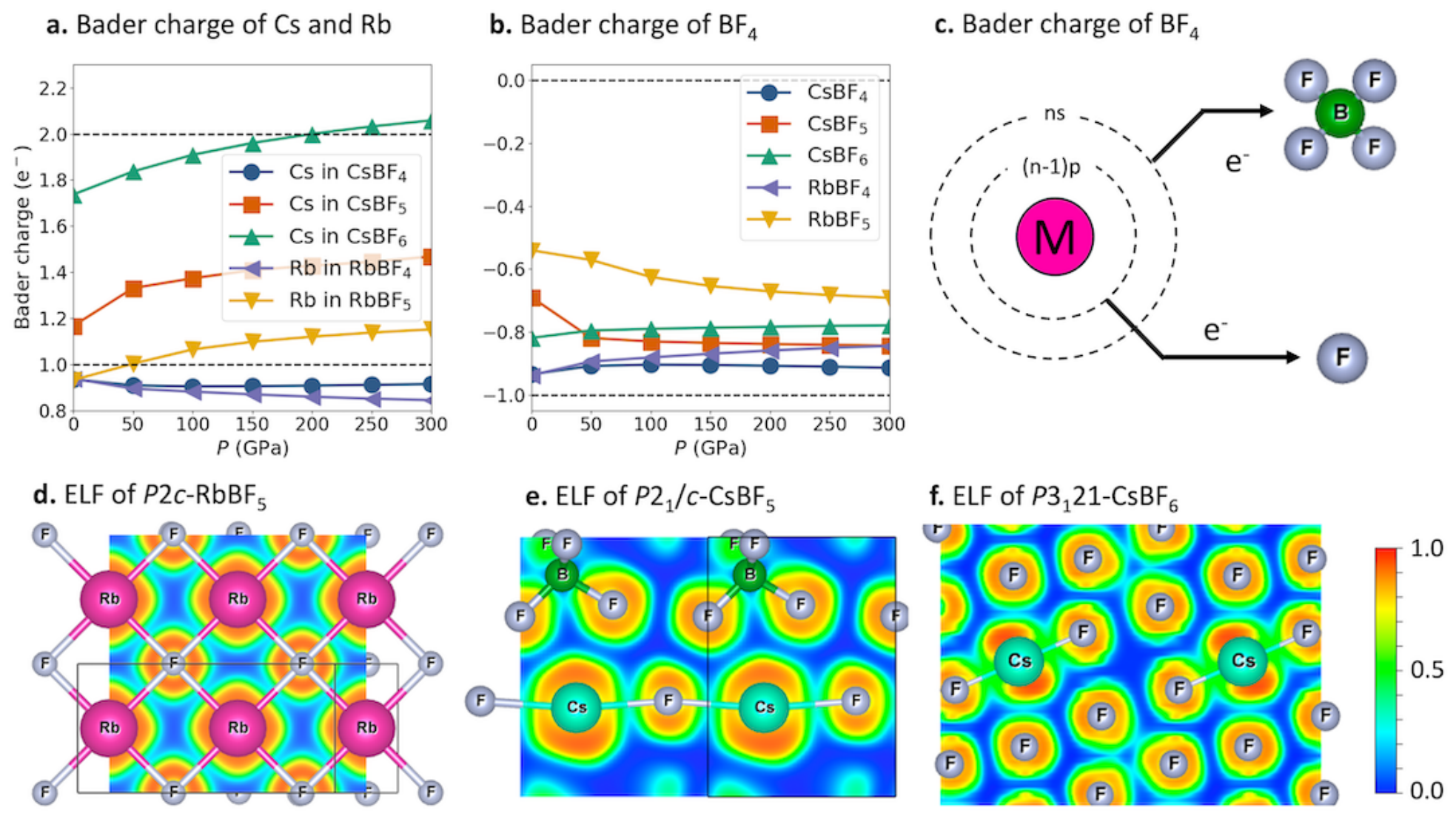

**Figure 3. The Bader charge and ELF**. **a.** The Bader charge of Rb and Cs in the synthesizable ternaries between 0 GPa and 300 GPa. **b.** The Bader charge of BF4 molecular units in the ternary systesm between 0 GPa and 300 GPa. **c.** the diagram of electron transferring related the n*s* and (n-1)*p* orbital in alkali metal Cs or Rb, denated as M here. **d-f** are the ELFs of *P*2/*c*-RbBF5 at 250 GPa, *P*2$_1$/*c*-CsBF5 at 200 GPa and *P*3$_1$21-CsBF6 at 200 GPa, respectively.

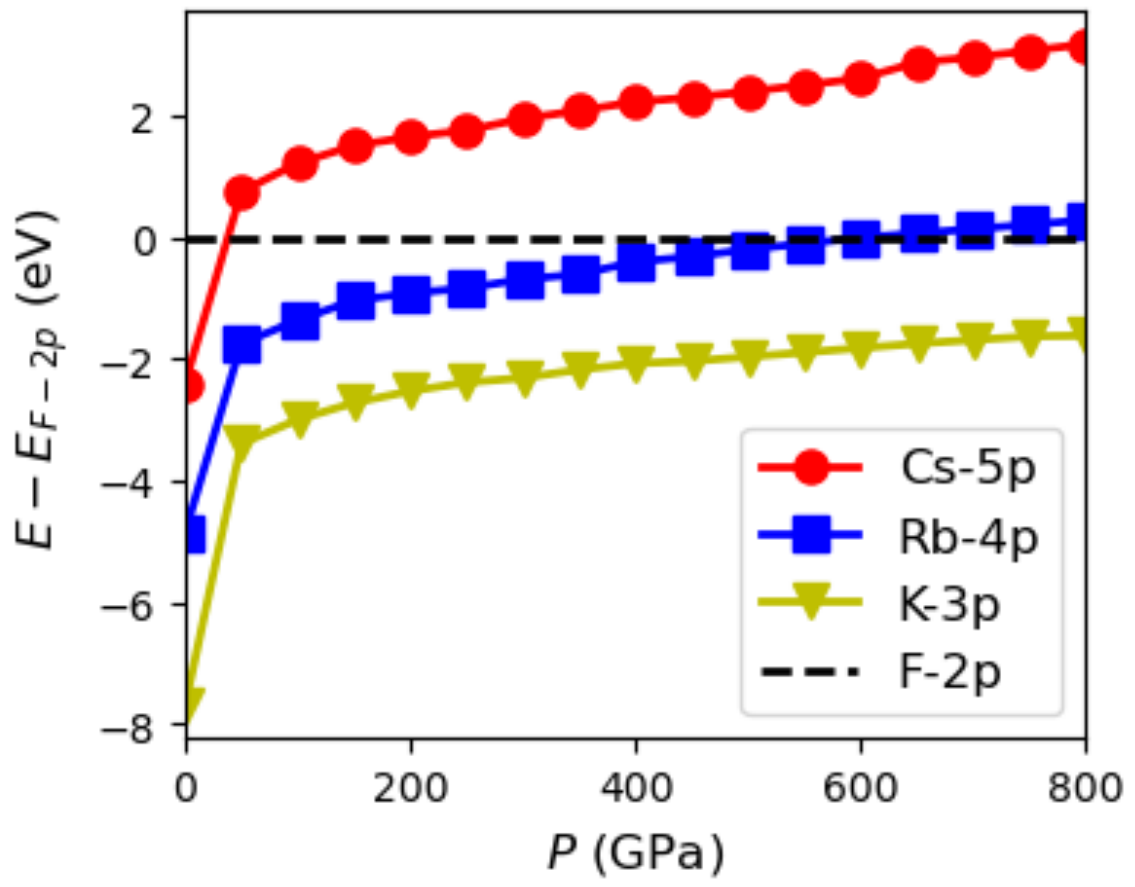

**Figure 4. The semi-core p orbtia level under pressure.** The semi-core *p*-orbital level of Cs, Rb, K relative to F-2*p* are computed between 0 and 800 GPa. The pressure is modeled by replacing one of the He atom in $3 \times 3 \times 3$ its f.c.c lattice with the target atom.

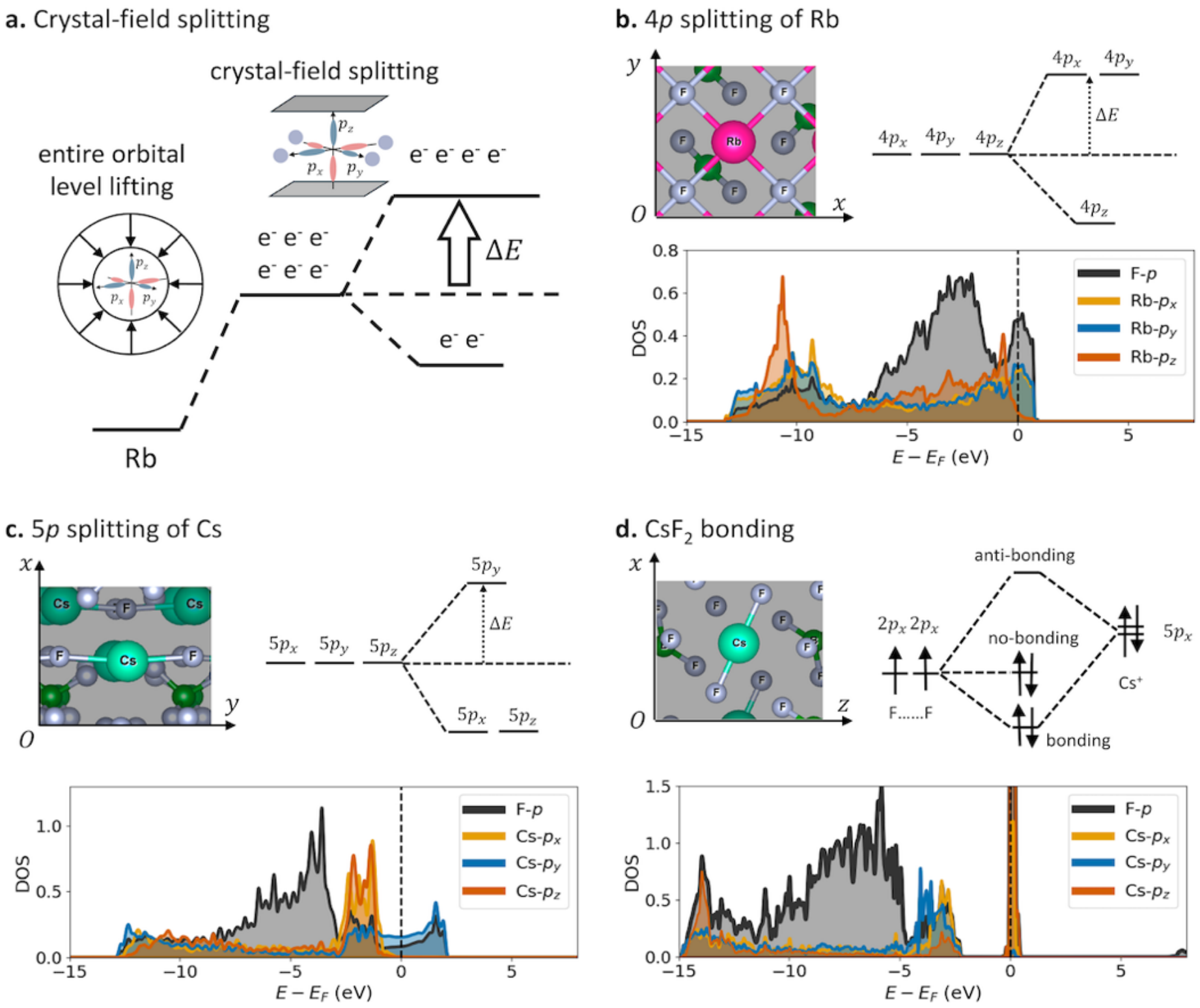

**Figure 5. Atomic orbital level and crystal field splitting of inner p-orbital of Rb and Cs. a.** Diagram illustrating the energy raising by crystal-field. **b** and **c.** the asymmetric local coordinates environment of Rb in P2/c-RbBF5 and Cs in P21c-CsBF5, respectively, as well as the inner p-orbital splitting in the scope of projected density of states (PDOS) of electron. **d.** the CsF2 bonding diagram and PDOS

Table 1. The ICOHP in *P*2/*c*-$RbBF_5$, *P*2₁/*c*-$CsBF_5$ and *P*3₁21-$CsBF_6$

| Fomula | Space group | Pair | Bond length(Å) | Pressure (GPa) | ICOHP (eV per pair) |
|---|---|---|---|---|---|
| $RbBF_5$ | *P*2/*c* | Rb-F | 2.055 | 250 | -0.930 |
| $CsBF_5$ | *P*$2_1$/*c* | Cs-F | 2.201 | 100 | -1.496 |
| | | Cs-F | 2.114 | 100 | -1.236 |
| $CsBF_6$ | *P*$3_1$21 | Cs-F | 1.853 | 200 | -3.844 |

## Supplementary Information

# Pressure-Driven Activation of Semi-core *p*-Orbitals in Rb and Cs via Crystal-Field Splitting

Shuran Ma,[1] Shuyang Sun,[1] Dengyao Yu,[1] Xue Cong,[1] Yanchao Wang,[2]* Yuanzheng Chen,[3]* and Zhen Liu[1,4]*

[1] *School of Physics and Astronomy, Beijing Normal University, Beijing 100875, China*

[2] *State Key Laboratory of Superhard Materials and International Center of Computational Method & Software, College of Physics, Jilin University, Changchun 130012, China*

[3] *School of Physical Science and Technology, Southwest Jiaotong University, Chengdu 610031, China*

[4] *Key Laboratory of Multiscale Spin Physics (Ministry of Education), Beijing Normal University, Beijing 100875, China*

**Table S1.** The calculated lattice parameters and atomic positions for the selected $CsBF_4$ structures.

| Formula | Space group | Pressure (GPa) | Lattice parameters (Å) | | | Fractional coordinates | | | |
|---|---|---|---|---|---|---|---|---|---|
| $CsBF_4$ | $Pna2_1$ | 50 | $a$=4.0770 | $b$=11.6638 | $c$=4.9580 | Cs | 0.4657 | 0.1023 | 0.7180 |
| | | | $\alpha$=90.0000 | $\beta$=90.0000 | $\gamma$ =90.0000 | B | -0.0987 | 0.1409 | 0.2183 |
| | | | | | | F | 0.5636 | 0.1287 | 0.2180 |
| | | | | | | F | -0.0092 | 0.1954 | -0.0138 |
| | | | | | | F | 0.5096 | 0.6956 | -0.0498 |
| | | | | | | F | 0.0420 | 0.0331 | 0.2187 |
| $CsBF_4$ | $Pnma$ | 300 | $a$=6.2795 | $b$=4.0833 | $c$=5.7684 | Cs | 0.3773 | 0.2500 | 0.5335 |
| | | | $\alpha$=90.0000 | $\beta$=90.0000 | $\gamma$ =90.0000 | B | 0.3439 | 0.2500 | 0.1004 |
| | | | | | | F | 0.4407 | -0.0150 | 0.1540 |
| | | | | | | F | 0.7656 | 0.2500 | 0.6079 |
| | | | | | | F | 0.1695 | 0.2500 | 0.2284 |

**Table S2.** The calculated lattice parameters and atomic positions for the selected $CsBF_5$ structures.

| Formula | Space group | Pressure (GPa) | Lattice parameters (Å) | | | Fractional coordinates | | | |
|---|---|---|---|---|---|---|---|---|---|
| $CsBF_5$ | $P2_1/c$ | 100 | $a$=5.9767 | $b$=4.3022 | $c$=8.5568 | Cs | 0.3423 | 0.2985 | 0.3713 |
| | | | $\alpha$=90.0000 | $\beta$=92.6975 | $\gamma$ =90.0000 | B | 0.8539 | 0.2226 | 0.3918 |
| | | | | | | F | -0.0281 | 0.3635 | 0.7705 |
| | | | | | | F | -0.0280 | 0.2478 | 0.5267 |
| | | | | | | F | 0.7386 | 0.4930 | 0.3675 |
| | | | | | | F | 0.6938 | 0.5004 | 0.8985 |
| | | | | | | F | 0.3679 | 0.7883 | 0.3648 |
| $CsBF_5$ | $P2_12_12$ | 100 | $a$=4.2988 | $b$=4.3071 | $c$=5.9663 | Cs | 0.0000 | 0.5000 | 0.8485 |
| | | | $\alpha$=90.0000 | $\beta$=90.0000 | $\gamma$ =90.0000 | B | 0.0000 | 0.5000 | 0.3581 |
| | | | | | | F | 0.2388 | 0.0062 | 0.5247 |
| | | | | | | F | 0.5019 | 0.2511 | 0.7797 |
| | | | | | | F | 0.0000 | 0.0000 | 0.8711 |

**Table S3.** The calculated lattice parameters and atomic positions for the selected $CsBF_6$ structures.

| Formula | Space group | Pressure (GPa) | Lattice parameters (Å) | | | Fractional coordinates | | | |
|---|---|---|---|---|---|---|---|---|---|
| $CsBF_6$ | $P2_13$ | 50 | $a$=6.5889 | $b$=6.5889 | $c$=6.5889 | Cs | 0.0118 | 0.0118 | 0.0118 |
| | | | $\alpha$=90.0000 | $\beta$=90.0000 | $\gamma$ =90.0000 | B | 0.4825 | 0.4825 | 0.4825 |
| | | | | | | F | 0.3640 | 0.6201 | 0.5884 |
| | | | | | | F | 0.8153 | 0.8153 | 0.8153 |
| | | | | | | F | 0.6840 | 0.6840 | 0.6840 |
| | | | | | | F | 0.3634 | 0.3634 | 0.3634 |
| $CsBF_6$ | $P3_121$ | 200 | $a$=3.3564 | $b$=3.3564 | $c$=14.7540 | Cs | 0.3670 | 0.0000 | 0.3333 |
| | | | $\alpha$=90.0000 | $\beta$=90.0000 | $\gamma$ =120.0000 | B | 0.0225 | 0.0000 | 0.8333 |
| | | | | | | F | 0.6499 | 0.5728 | 0.2818 |
| | | | | | | F | 0.7270 | 0.7368 | 0.4300 |
| | | | | | | F | 0.6645 | 0.6053 | 0.1508 |

**Table S4.** The calculated lattice parameters and atomic positions for the selected $RbBF_4$ structures.

| Formula | Space group | Pressure (GPa) | Lattice parameters (Å) | | | Fractional coordinates | | | |
|---|---|---|---|---|---|---|---|---|---|
| $RbBF_4$ | $Pnma$ | 0 | $a = 5.7075$ | $b = 7.4100$ | $c = 9.2238$ | Rb | 0.2500 | 0.1612 | 0.8142 |
| | | | $\alpha = 90.0000$ | $\beta$ =90.0000 | $\gamma = 90.0000$ | B | 0.2500 | 0.6928 | -0.0621 |
| | | | | | | F | 0.5478 | 0.3037 | 0.5769 |
| | | | | | | F | 0.2500 | 0.6130 | 0.0767 |
| | | | | | | F | 0.2500 | 0.5607 | 0.8285 |
| $RbBF_4$ | $P4/n$ | 200 | $a = 5.3238$ | $b = 5.3238$ | $c = 2.6696$ | Rb | 0.2500 | 0.2500 | 0.2551 |
| | | | $\alpha = 90.0000$ | $\beta = 90.0000$ | $\gamma = 90.0000$ | B | 0.2500 | 0.7500 | 0.0000 |
| | | | | | | F | -0.069 | 0.3628 | 0.7467 |

**Table S5.** The calculated lattice parameters and atomic positions for the selected $RbBF_5$ structures.

| Formula | Space group | Pressure (GPa) | Lattice parameters (Å) | | | Fractional coordinates | | | |
|---|---|---|---|---|---|---|---|---|---|
| $RbBF_5$ | $P2_12_12_1$ | 0 | $a$ = 13.1850 | $b$ = 4.4490 | $c$ = 7.6965 | Rb | 0.3394 | 0.7162 | 0.0781 |
| | | | $\alpha$ = 90.0000 | $\beta$ =90.0000 | $\gamma$ = 90.0000 | B | 0.4091 | -0.035 | 0.5038 |
| | | | | | | F | 0.0298 | 0.8745 | 0.1206 |
| | | | | | | F | 0.5296 | 0.2750 | 0.0975 |
| | | | | | | F | 0.3568 | 0.1673 | 0.3932 |
| | | | | | | F | 0.6941 | 0.7238 | 0.6429 |
| | | | | | | F | 0.8361 | 0.7035 | 0.4013 |
| $RbBF_5$ | $P2/c$ | 200 | $a$ = 5.9151 | $b$ = 2.9763 | $c$ = 5.0437 | Rb | 0.0000 | 0.5000 | 0.5000 |
| | | | $\alpha$ = 90.0000 | $\beta$ = 106.5459 | $\gamma$ = 90.0000 | B | 0.2500 | 0.2397 | 0.0000 |
| | | | | | | F | 0.2500 | 0.9975 | 0.5000 |
| | | | | | | F | 0.3365 | 0.5041 | 0.8433 |
| | | | | | | F | 0.9189 | 0.0047 | 0.1578 |

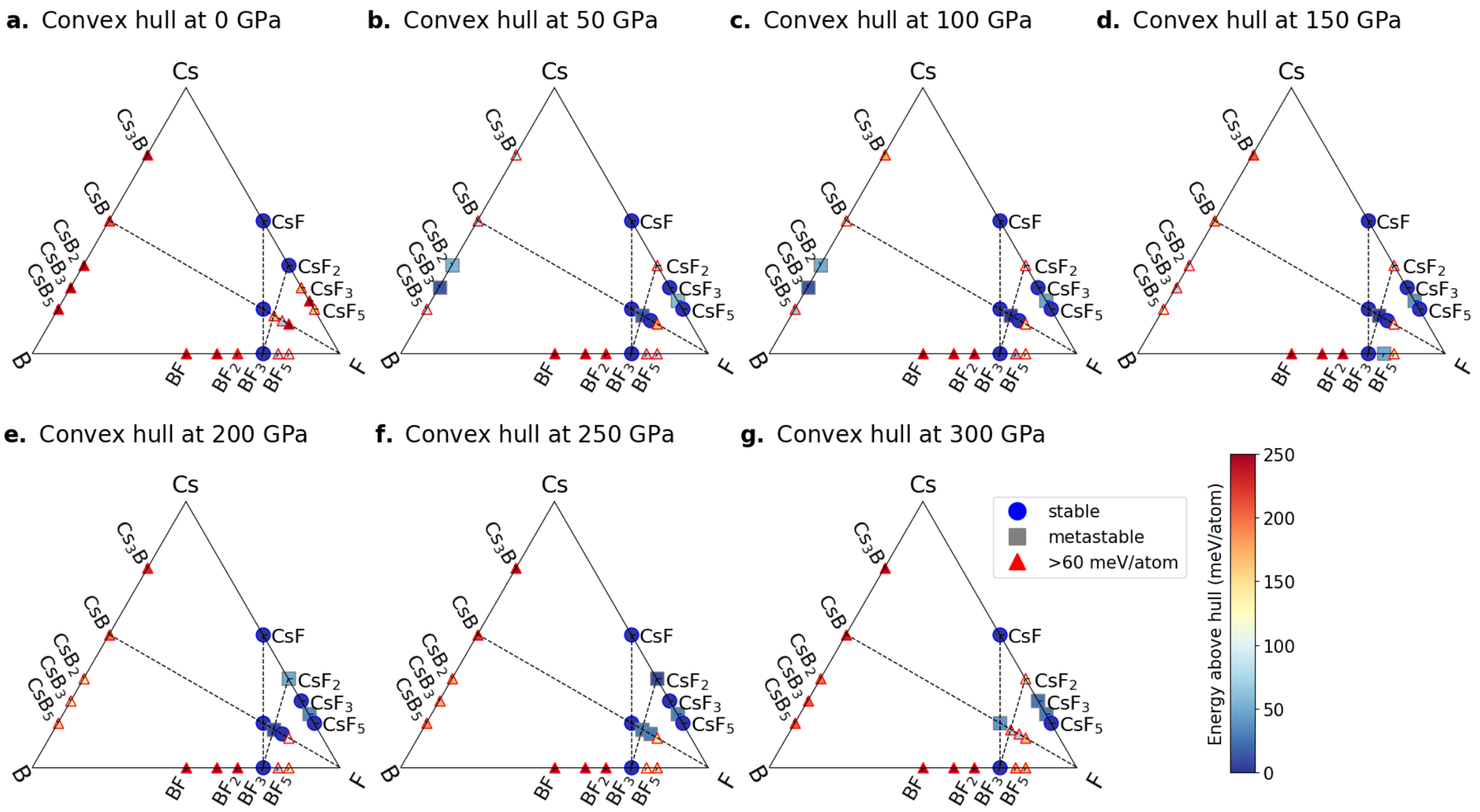

**Figure S1.** The triangle diagram in terms of the value of the formation enthalpy above convex hull at 0GPa to 300 GPa. The stoichiometries are classified into three categories, whics are stable, metastable and unstable. The stable stoichiometries refer as the ones whose formation enthalpy lying on the convex hull. The formation enthalpy of the metastable is higher than the convex hull with 60 emV per atom.

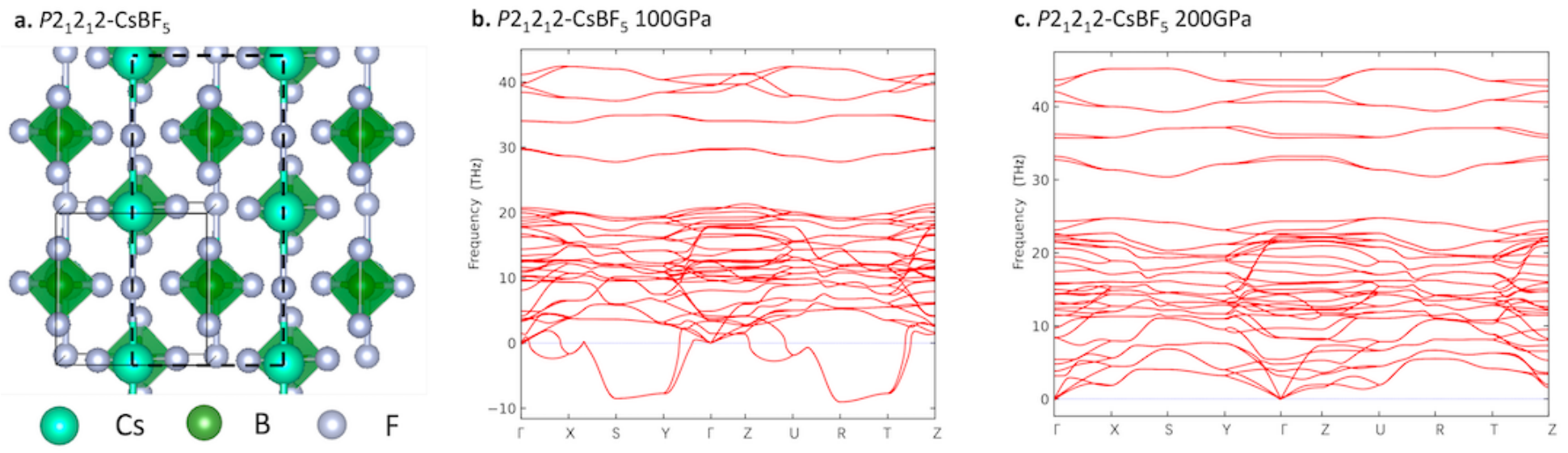

**Figure S2**. The structure of $P2_12_12$-$CsBF_5$ is in **a** and its phonon dispersion relation at b**.** 100 GPa and c**.** 200 GPa. The phonon dispersion relation at 100 GPa suggest dynamical instability for this structure.

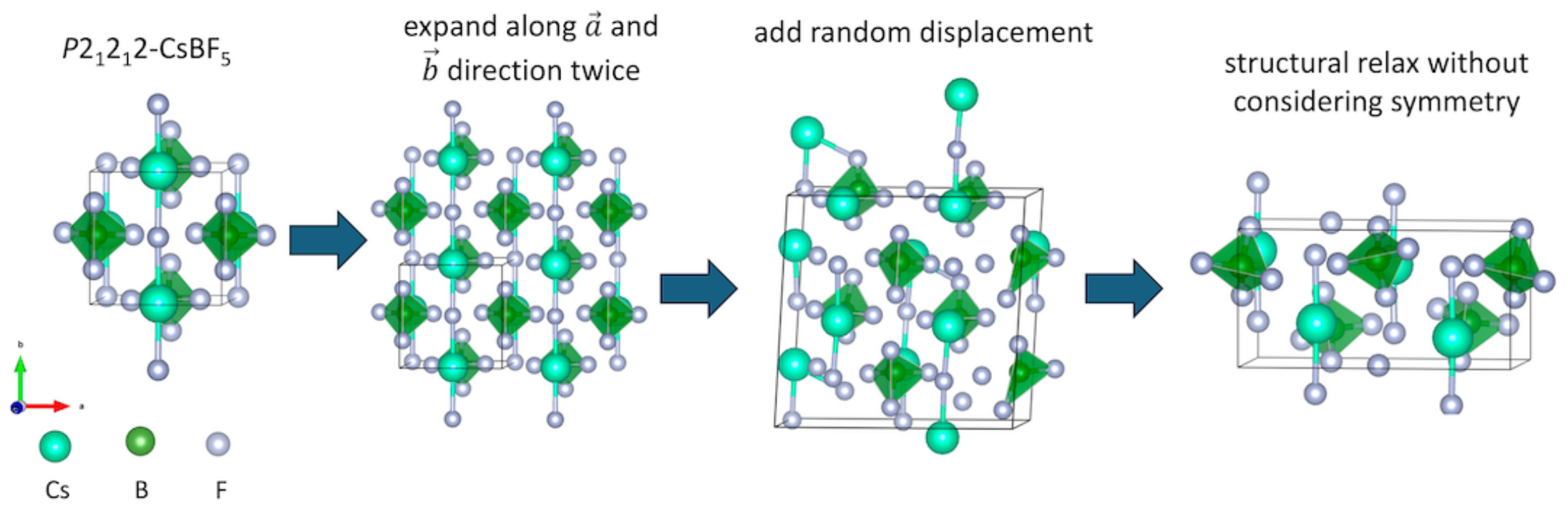

**Figure S3**. Diagram for relaxation from a structure by add random displacement to $P2_12_12$-$CsBF_5$. In Figure S2, the phonon dispersion relation at 100 GPa suggest dynamical instability. We pick the S (0.5, 0.5, 0.0) point in the Brilliuon zone for an attempt for search new structure with a relaxation. The twice expansion along a and b is commensurate with S (0.5, 0.5, 0.0) point. With adding random displacement on the supercell, a structure relaxation without considering the symmetry (set ISYM = 0 in INCAR in VASP computation) generates a new structure for $CsBF_5$, whose space group is $P2_1/c$.

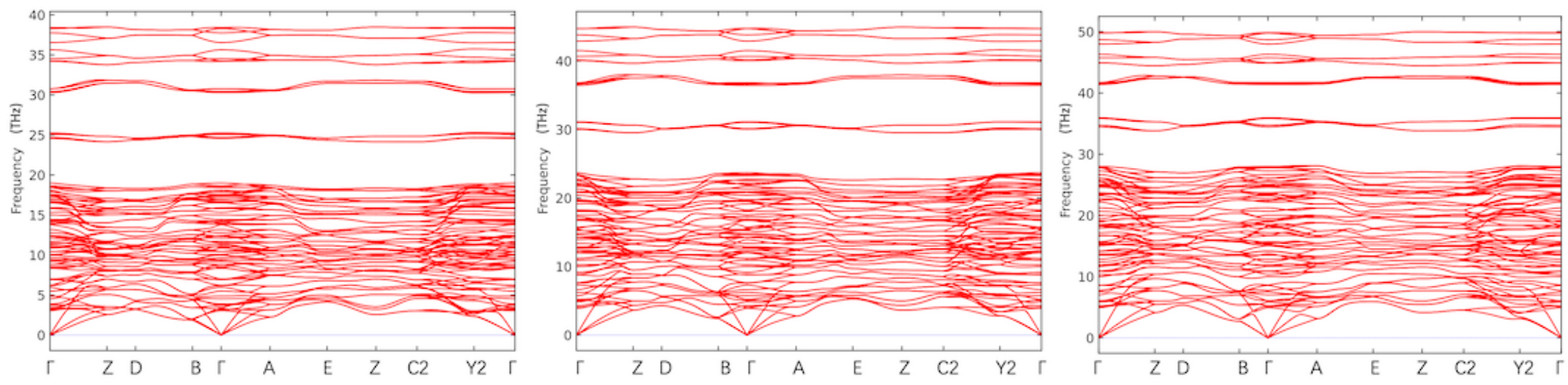

**Figure S4.** $P2_1/c$-$CsBF_5$ phonon dispersion relation at **a.** 100 GPa, **b.** 200 GPa and **c.** 300 GPa. This structure is dynamically stable in a wide range of pressure.

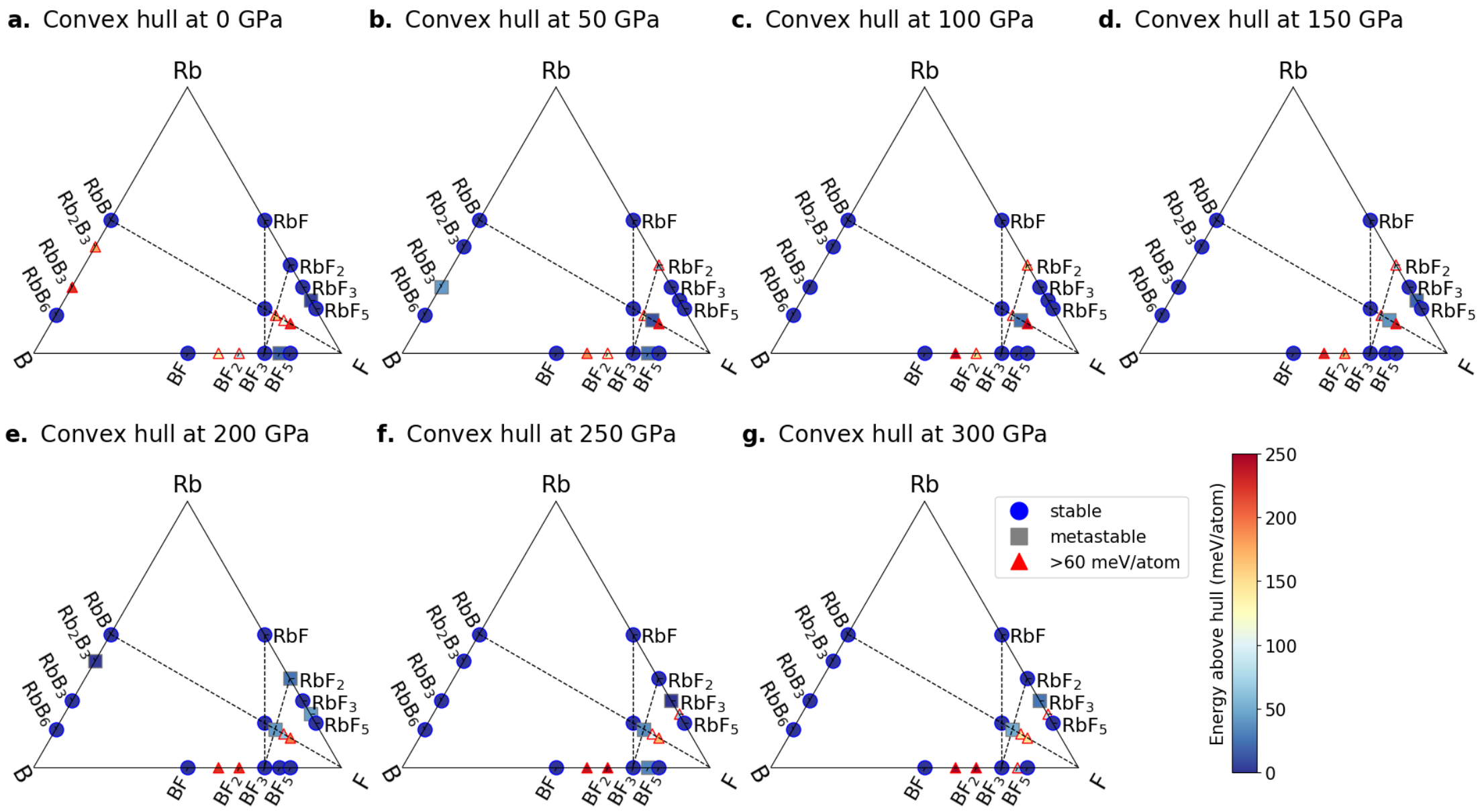

**Figure S5.** The triangle diagram in terms of the value of the formation enthalpy above convex hull at **a.** 0 GPa, **b.** 100 GPa, and **c.** 200 GPa, and **d.** 300 GPa, respectively. The stoichiometries are classified into three categories, whics are stable, metastable and unstable. The stable stoichiometries refer as the ones whose formation enthalpy lying on the convex hull. The formation enthalpy of the metastable is higher than the convex hull with 60 emV per atom.

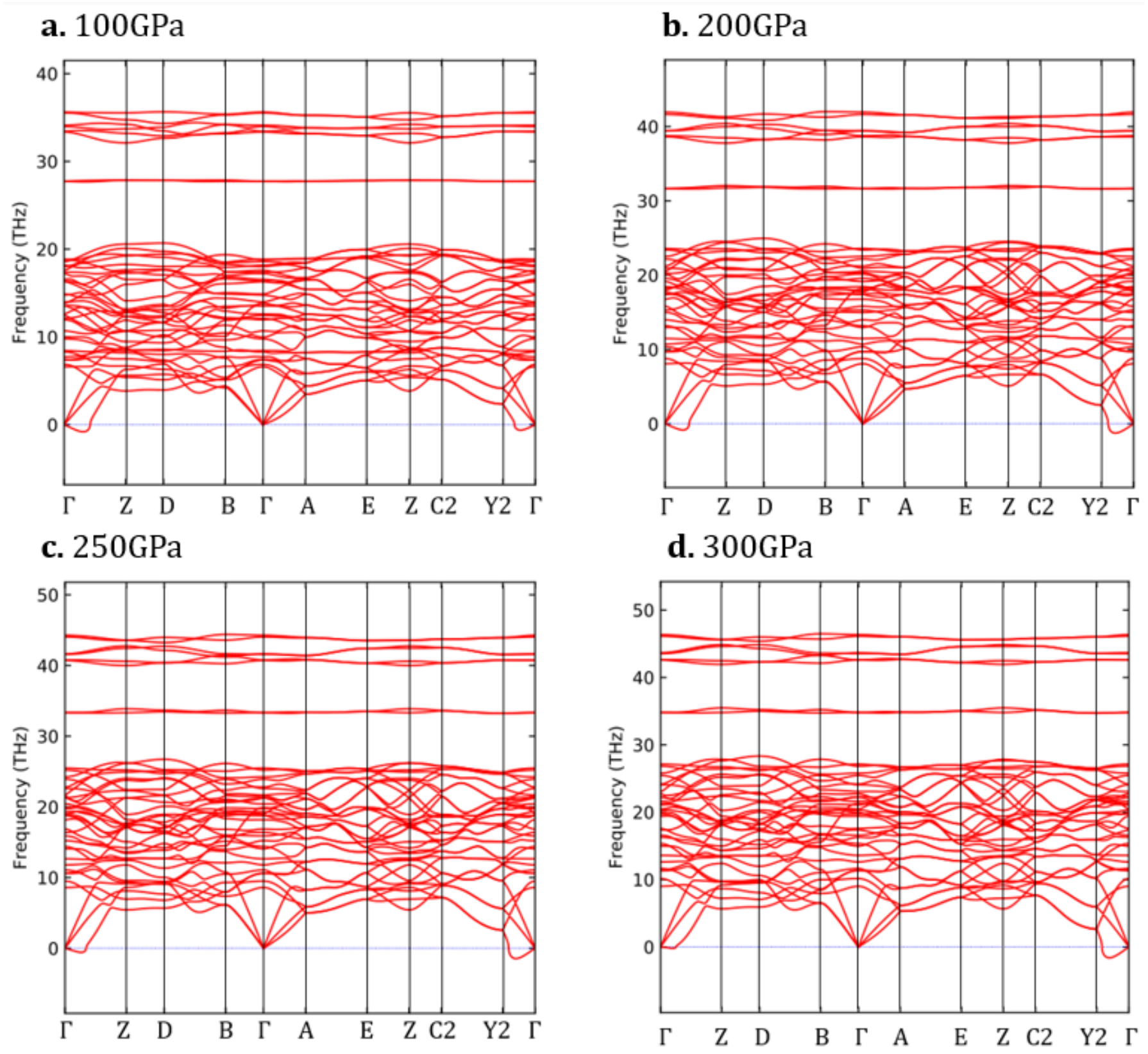

**Figure S6.** The phonon band of $P2/c$-$RbBF_5$ at **a.** 100 GPa. **b.** 200 GPa, **c.** 250 GPa and **d.** 300 GPa. Under these four pressures, the imaginary frequencies are all small, no more than 0.5 THz and 2.0 THz along the paths of Γ-Z and Γ-Y2, respectively. We use the similar process mentioned in Figure S3, but the structure with random displacement always turn back to the original $P2/c$-$RbBF_5$, which is an indicator that this structure resides at the local minimum value of potential energy surface.

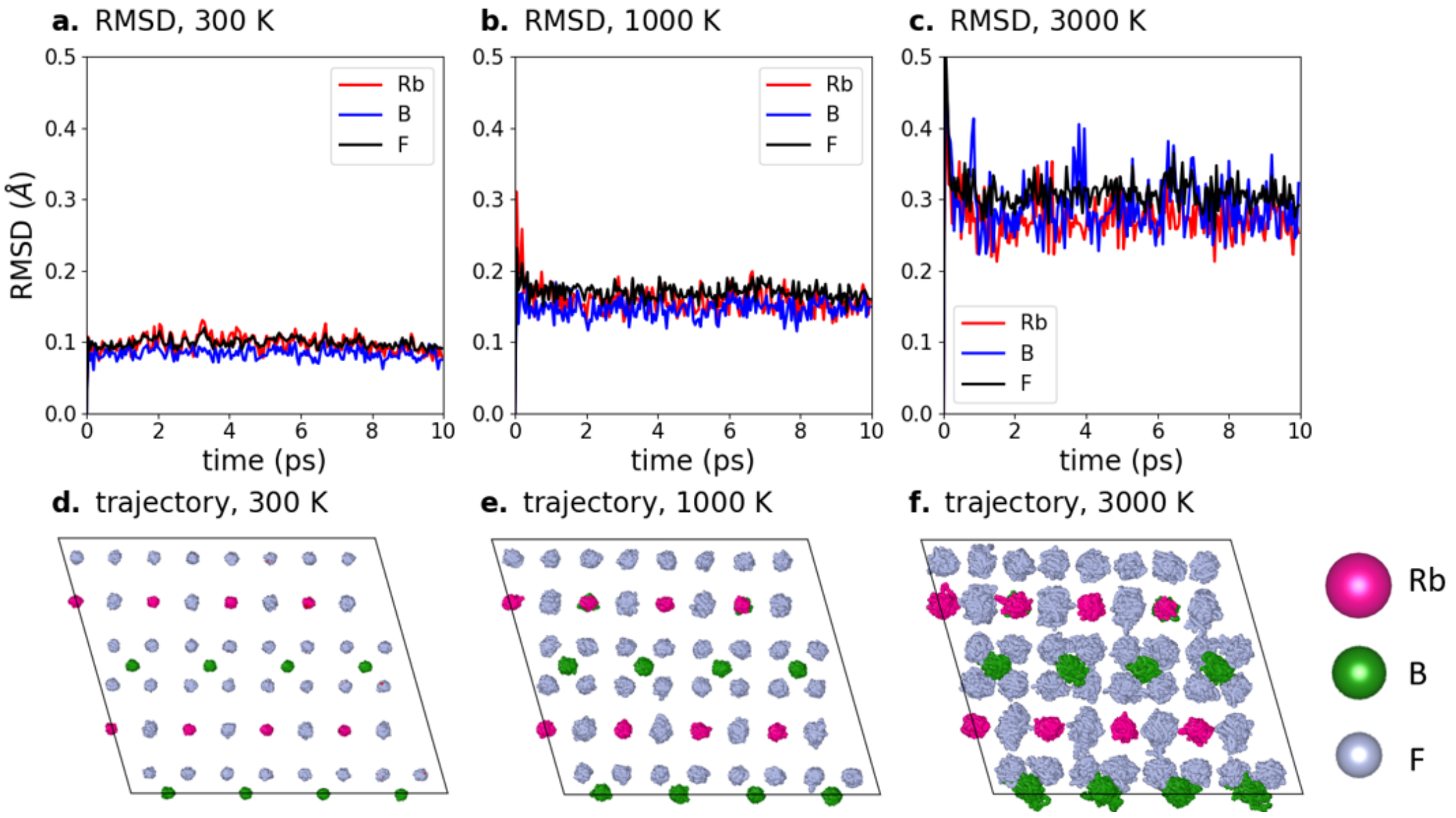

**Figure S7.** The RMSD and trajectories of $P2/c$-$RbBF_5$ at 100 GPa. **a-c**. The RMSD at 300 K, 1000 K and 3000 K, respectively. **d-f**. The trajectoies at 300 K, 1000 K and 3000 K, resspectively.

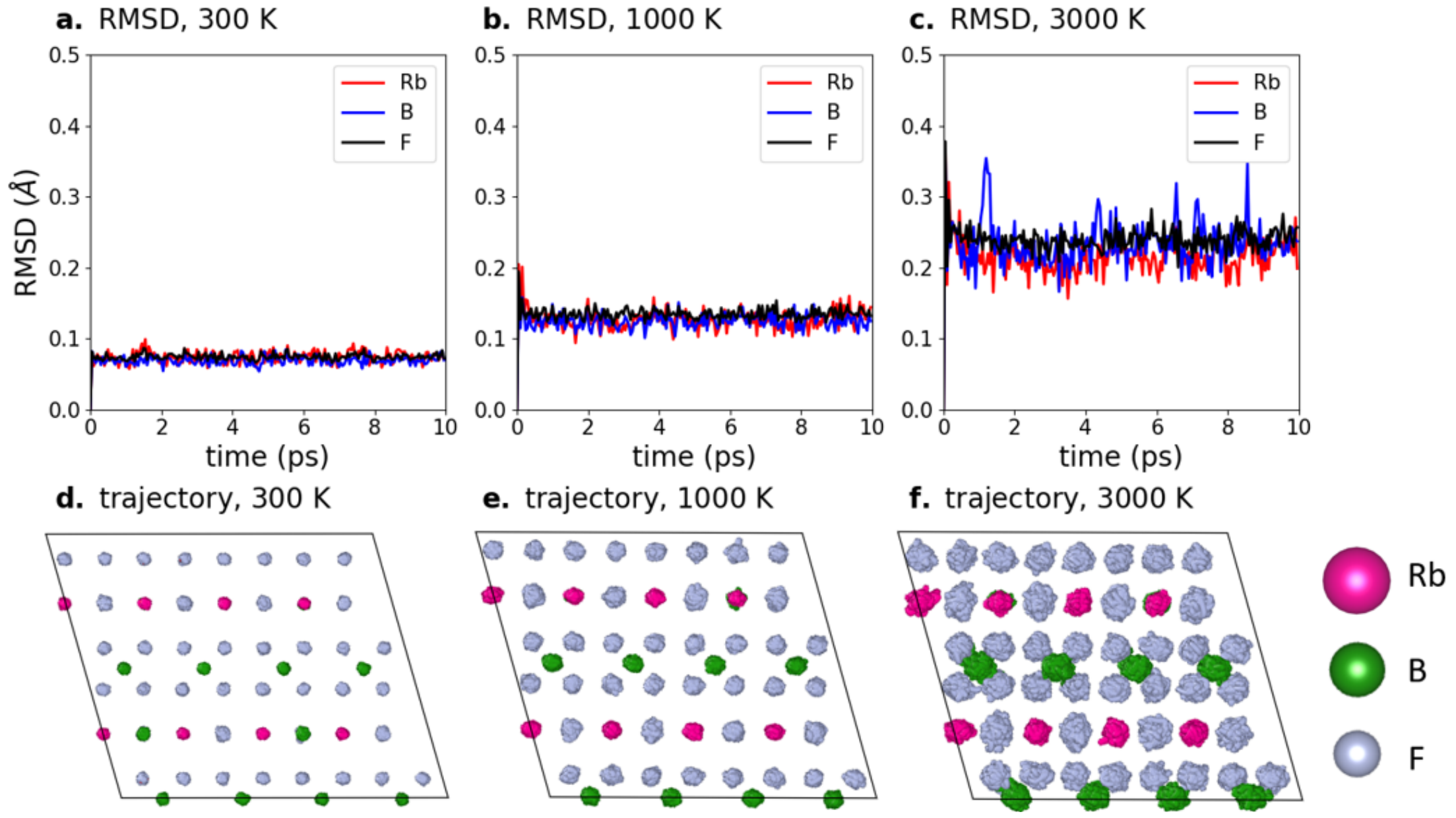

**Figure S8.** The RMSD and trajectories of $P2/c$-$RbBF_5$ at 200 GPa. **a-c**. The RMSD at 300 K, 1000 K and 3000 K, respectively. **d-f**. The trajectoies at 300 K, 1000 K and 3000 K, resspectively.

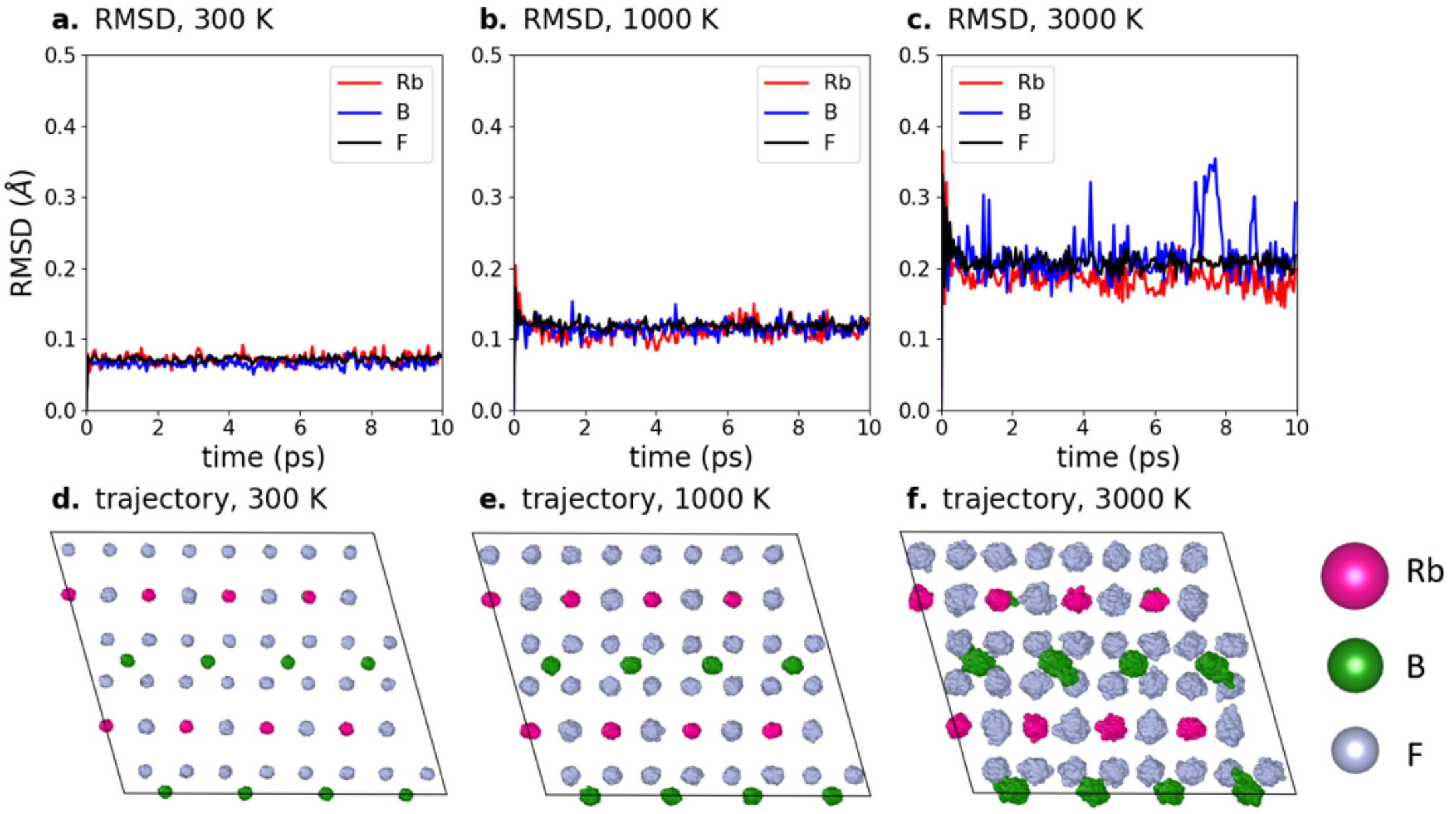

**Figure S9.** The RMSD and trajectories of $P2/c$-$RbBF_5$ at 300 GPa. **a-c**. The RMSD at 300 K, 1000 K and 3000 K, respectively. **d-f**. The trajectoies at 300 K, 1000 K and 3000 K, resspectively.